\documentclass[twoside]{article}
\usepackage{fleqn,espcrc2}

\usepackage{graphicx}
\usepackage[figuresright]{rotating}
\usepackage{epsfig}

\newcommand{\AmS}{{\protect\the\textfont2
  A\kern-.1667em\lower.5ex\hbox{M}\kern-.125emS}}

\title{TAUOLA the library for $\tau$ lepton decay,
 and KKMC/KORALB/KORALZ/... 
 status report }

\author{ Z. W\c{a}s\address[MCSD]{Institute of Nuclear Physics,\\
         ul. Kawiory 26a, 30-055 Cracow, Poland}%
        \thanks{Supported in part by Polish Government grant 
KBN 2P03B11819, 
the Maria Sk\l{}odowska-Curie Joint Fund II PAA/DOE-97-316,
 European Comission 5-th Framework grant HPRN-CT-2000-00149,
and the Polish--French Collaboration within IN2P3 through LAPP Annecy;
Home page at http://wasm.home.cern.ch/wasm/},
}
       
\begin{document}

\begin{abstract}
The status of the Monte Carlo programs for the simulation of the $\tau$ lepton production in 
high energy accelerator experiments and decay is reviewed. In particular, the status of the
following packages is discussed: (i) {\tt TAUOLA} for $\tau$-lepton decay, (ii) {\tt PHOTOS} for 
radiative corrections in decays,  (iii) {\tt KORALB, KORALZ, KKMC} packages for $\tau$-pair production 
in $e^+e^-$ collisions and (iv)  universal
interface of {\tt TAUOLA} for the decay of $\tau$-leptons produced by``any'' generator. 
Special emphasis on requirements from new and future experiments is
given. Some considerations about the software organization necessary to 
keep simultaneously distinct physics initializations for {\tt TAUOLA} 
are also included. 
\vspace{1pc}
\end{abstract}

\maketitle

\section{Introduction}
The {\tt TAUOLA} package
\cite{Jadach:1990mz,Jezabek:1991qp,Jadach:1993hs,Golonka:2000iu}  for the simulation 
of $\tau$ lepton decays and  
{\tt PHOTOS} \cite{Barberio:1990ms,Barberio:1994qi} for the simulation of radiative corrections
in decay, are computing
projects with a rather long history. Written and maintained by 
well defined authors, they nonetheless migrated into a wide range
of applications where they became ingredients of 
complicated simulation chains. As a consequence, a large number of
different versions are presently in use.
From the algorithmic point of view, they often
differ only in a few small details, but incorporate a
substantial amount of specific results from the distinct
$\tau$-lepton measurements. Such versions were mainly maintained (and will remain so maintained) 
by the experiments taking precision data on $\tau$ leptons. On the other hand,
 many new applications were developed  recently,  often requiring
 program versions which differ because of
interfaces to other packages  (eg. the format
of the event record had to be adjusted, the organization of the input parameters 
changed, etc.).

In the last two years, substantial progress for  the simulation 
of $\tau$-pair production was achieved. Control of the
systematic errors, for the theoretical predictions, at the few permille level was achieved 
(or, for some centre-fo-mass system energies, is relatively easy to achieve in the near future)
for the centre of mass energy from the $\tau$-pair production threshold up to the
energies of the future linear colliders. 

\section{Versions of TAUOLA Monte Carlo}

In ref.  \cite{Golonka:2000iu} the setup for constructing specific versions of {\tt TAUOLA}
and  {\tt PHOTOS} from the single set of files was prepared and documented. The system  
was  prepared for the software librarians and advanced users interested in updating
the packages for the multipurpose environment. 
The idea was to create a repository which allows one to  keep all main 
options of {\tt TAUOLA} developed for different purposes in a relatively compact form,
without duplications of semi-identical parts. The
repository was set to produce the standard {\tt FORTRAN} files 
which can be later handled, 
exactly the same way, as the published versions of the packages.

\vskip 3 mm
{\bf Motivations for versioning}:

\begin{enumerate}
\item
{\tt PHOTOS}: Versions of the {\tt FORTRAN} code are necessary because of the different versions of
        the {\tt HEPEVT} common block being in use in the HEP libraries 
        (single/double precision,  dimensionality of storage matrices). 
\item
{\tt TAUOLA}: Versions of the {\tt FORTRAN} code are motivated by: (A) different versions of
        initialization of physics parameters; (B)
        interfaces to different Monte Carlo generators for production of $\tau$-lepton(s); 
        and
        (C) different versions of the {\tt HEPEVT} common block:
        \begin{itemize}
        \item
(A) Different physics initializations: \\
(1) As published in \cite{Jadach:1993hs}; \\
(2) As initialized by the ALEPH collaboration  \cite{aleph} (it is suggested to use this 
    version only with the help of the collaboration's advice); \\
(3) As initialized by the CLEO collaboration \cite{cleo} (see the printout of this 
   version for details); \\
(4) Further coding of some individual decay modes.
\item
(B) Different interfaces with MC generators:\\
 (1) Old demo program as in the published version \cite{Jadach:1993hs}; \\
 (2) Interface to {\tt KKMC} \cite{kkcpc:1999}; \\
 (3) New universal interface using {\tt HEPEVT}  common block.
\item
(C) Different versions of the {\tt HEPEVT} common block.\\
\end{itemize}
\item
 TAUOLA and PHOTOS: different versions of random number generators. 
\item
 TAUOLA and PHOTOS: makefiles with different compiler flags.
\end{enumerate}

Standard tools are used in the discussed setup:
{\tt cpp} the C-language pre-compiler: its {\tt if}, {\tt elif} and {\tt include} commands, as well
as {\tt UNIX} logical links and {\tt cat} command.
The aim is to provide full backward compatibility at the level of the  {\tt FORTRAN}
source with the published  versions, or other versions 
being in use at present.  
It is expected that one will use the setup to create her/his version of the {\tt TAUOLA}
and {\tt PHOTOS} libraries and interfaces  (subdirectories {\tt tauola/} and {\tt
photos/}) consisting of the plain {\tt FORTRAN} code. The original  subdirectories 
of the setup can then be erased or stored for future use.

\vskip 2mm
\centerline{\bf The following directories are created}
\centerline{\bf once the actions of the setup are completed:}
\vskip 2mm

\begin{enumerate}
  \item
  {\tt photos/}: Standard directory with the {\tt FORTRAN} code of the {\tt PHOTOS}
  library and its demo.
  \item
  {\tt tauola/}: Standard directory with the {\tt FORTRAN} code of the {\tt TAUOLA} library, 
                    its demos and example outputs. 
  \begin{enumerate}
    \item
    {\tt tauola/demo-standalone}: Demo program for {\tt TAUOLA} executed in a standalone mode.
    \item
    {\tt tauola/demo-jetset}: Demo program for {\tt  TAUOLA} executed with
    universal interface to physics event generators
    based on the
    {\tt HEPEVT} common block. In this demo {\tt HEPEVT} is filled 
    from the {\tt JETSET74} \cite{jetset6.3:1987} Monte Carlo generator.
   \item
   {\tt tauola/demo-KK-face}: Interface to the KK Monte Carlo \cite{kkcpc:1999}. 
  \end{enumerate}
\end{enumerate}

\subsection{Options for PHOTOS Monte Carlo}
The different options of {\tt PHOTOS} which can be created\footnote{ Executing
single command: {\tt make xxx-all}.}
correspond solely to
the different versions of the {\tt HEPEVT} common block.
The possible options are:
\begin{enumerate}
\item
{\tt KK-all}     -- for KK Monte Carlo
\item
{\tt 2kD-all}    -- dimension  2000 double precision
\item
{\tt 4kD-all}    -- dimension  4000 double precision
\item
{\tt 2kR-all}    -- dimension  2000 single precision
\item
{\tt 10kD-all}   -- dimension 10000 double precision
\end{enumerate}

\subsection{Options for TAUOLA Monte Carlo}

Basic options for the physics initializations, activated with the single command 
{\tt make},  are: {\tt cpc}; {\tt cleo};
{\tt aleph}. The three possible versions of the created {\tt tauola} directory correspond to
form-factors and branching ratios defined respectively as in: 
({\tt cpc}) the published version of {\tt TAUOLA}; ({\tt aleph}) the conventions as adopted by the
ALEPH collaboration, 
({\tt cleo}) the conventions as adopted by the CLEO collaboration.

   $\bullet$ Additional parametrizations for form-factors, which can be useful
   in some applications, are placed in the separate directory. At present,
   the code used in refs. \cite{Abbiendi:1999cq} and \cite{Abreu:1998cn} 
   is stored there.

   $\bullet$ 
It is often necessary to change some of the {\tt TAUOLA}  input
parameters like branching
ratios, the mass of the $\tau$-lepton, etc. It is convenient to have it done once for 
all applications i.e. {\tt demo-KK-face}, {\tt demo-jetset} and {\tt demo-standalone}.
The special arrangements to do it, in a consistent way, are provided.
For details of the initialization routines, which are semi-identical in the three cases, 
see refs. 
\cite{Jadach:1990mz,Jezabek:1991qp,Jadach:1993hs}.
Note that special care from the physics point of view is needed. Often,
input parameters  are inter-related with the actual choice 
of form factors. The changes should be thus performed simultaneously.

\section{Universal interface based on {\tt HEPEVT} common block}

The universal interface of {\tt TAUOLA} for ``any'' $\tau$ production generator
is provided. It uses as an input, the {\tt HEPEVT} common block and operates on its content
only. As a demonstration example the
interface is combined  with the {\tt JETSET} generator, however it should work
in the same manner with the {\tt PYTHIA\footnote{It was already checked to be the case.}, 
HERWIG} or {\tt ISAJET} generators as well.

The interface work in the following way:
\begin{itemize}
\item
The $\tau$-lepton should be forced to be stable in the $\tau$ production  generator.
\item
The content of the {\tt HEPEVT} common block is searched for all $\tau$
leptons and neutrinos first. 
\item
It is checked if there are $\tau$-flavour pairs (two $\tau$-leptons or
$\tau$-lepton and $\tau$-neutrino) originating from the same mother. 
\item
The decays of the $\tau$-flavour pairs are performed with {\tt TAUOLA}.
Longitudinal spin effects are generated  in the case of the $\tau$ produced from decay of:
$W \to \tau \nu$, $Z/\gamma \to \tau \tau$, 
the neutral Higgs boson $H \to \tau \tau$, and the charged Higgs boson
 $H^{\pm}\to  \tau \nu$.
Parallel or anti-parallel spin configurations are
generated, before calling on the $\tau$ decay, and then the decays of 100 \% polarized $\tau$'s
are executed. 
\item
In the case of the Higgs boson (for the spin correlations to be generated)
the identifier of the $\tau$ mother must be that of the Higgs.
\item
In case of the $W$ and $Z/\gamma$ it is not necessary. 
If from the same mother as that of the $\tau$ there is  produced also
a $\nu_\tau$, the $W$ is reconstructed by the interface as the sum of the two. 
Simmilarily, $Z/\gamma$ is reconstructed if from the
same mother another $\tau$ is produced.

\item
Let us note that the calculation of the $\tau$ polarization created from
the $Z$ and/or virtual $\gamma$ (as function of the direction) represents 
a rather non-trivial extension.
Dedicated study of 
the production matrix elements of the host generator is necessary.
A separate paper \cite{Gosia} is devoted to this point; let us show however some 
preliminary results in the next section. At this moment the distribution version
of the package, available from  the www-page, does not include this possibility.
The updated version can be obtained from the authors upon individual requests only. 
\item
Photon radiation in decay is performed with {\tt PHOTOS}.
\end{itemize}

\section{Tau polarization from the $Z/\gamma$ mechanism}
The best way of calculating spin state of any final state is 
with the help of the matrix element and the rigorous density matrix treatment. 
This is however not always possible or necessary. Often, like in the
case of the production and decay of particles in the ultra-relativistic limit a 
simplified approach can be sufficient. Such an approach was developed 
for {\tt KORALZ} Monte Carlo program \cite{koralz4:1994} and its limitations were studied
with the help of  matrix element calculations of order $\alpha$ \cite{was:1987}.
In \cite{Gosia} we study the question of  whether a similar approach can not
be generalized, and the approximate spin correlation calculated from the 
information stored in the {\tt HEPEVT} common block by the typical $\tau$ production program:

The approximation consists of calculating/approximating variables and later using the
spin correlation of the elementary $2 \to 2$ body 
process $e^+ e^- (q\bar q) \to \tau^+\tau^- $, buried inside multi-body
production process. Let us stress that such a procedure can never be fully
controlled as its uncertainties and even functioning depends on the way the
particular production program fills the {\tt HEPEVT} common block. It will be 
always  responsibility of the user to check if in the particular case 
the approximation can be useful. Nonetheless our aim is {\it not} to replace
the matrix element calculations, but rather to provide a  method of calculating/estimating
spin effects in cases when otherwise spin effects would not be taken care of, at all.
Needless to say  such an approach is limited for the spin treatment to the
approximation not better than the leading-log.

The principle of calculating kinematical variables is simple. 
The 4-momenta of the  $2 \to 2$ body process have to be found.
The 4-momenta of the outcoming $\tau$'s are used 
directly. Initial state momenta are constructed from the incoming and outcoming 
momenta of the particles (or fields) accompanying production of the 
$Z/\gamma$ state\footnote{The $Z/\gamma$ state  does not need to be explicitly coded 
in the {\tt HEPEVT} common block. Note that if 
available, information from the history part of the event, where the 4-momenta
of gluons quarks etc. are stored, will be used.}. 
We group them accordingly to fermion number flow, and ambiguous additional particles
are grouped (summed) into effective quarks  to minimize their virtualities.
Such an approach is consistent in the case of emission of photons or gluons
with the leading log approximation.

In the following, let us give a few examples of tests we have performed so far on
the elements of the calculation, see 
ref \cite{Gosia} for more details.
In fig. 1 we plot the $\tau$ polarization in 
   $2 \to 2$ body scattering as a function of the scattering angle respectively for
$e^-e^+$, $u \bar u$, $d \bar d$ initial states. 
In fig. 2 we use the most spin sensitive decay mode $\tau \to \nu \pi$ and 
 we show the $\pi$ energy spectrum (in the $Z$ rest-frame), and 
energy-energy correlations between $\pi^+$ and  $\pi^-$ (also in the $Z$ rest-frame).
Due to the $Z$ interaction the $\tau$ leptons are polarized; there is thus a shift toward
lower energies in the $\pi$ energy spectrum. From the second part of the plot, we see
that the configurations where both $\pi^+$ and $\pi^-$ have energies bigger or smaller 
than average are more populous than the mixed configurations, exhibiting the 
vector nature of the $Z/\gamma-\tau-\tau$ vertex.
On the  contrary, in the case of the production from the Higgs meson, the opposite would be true,
and the mixed fast-slow (slow-fast) configurations would be more common, 
the energy spectrum of the $\pi$ would be flat.
\begin{figure}[!ht]
\centering
\setlength{\unitlength}{0.1mm}
\begin{picture}(800,1500)
\put( 350,1540){\makebox(0,0)[b]{ ($e^-e^+ \to \tau^+ \tau^-$)}}
\put( 350,975){\makebox(0,0)[b]{ ($u \bar{u}\to \tau^+ \tau^- $ )}}
\put( 350, 410){\makebox(0,0)[b]{ ( $d \bar{d}\to \tau^+ \tau^-$ )}}
\put(-20, 810){\makebox(0,0)[lb]{\epsfig{file=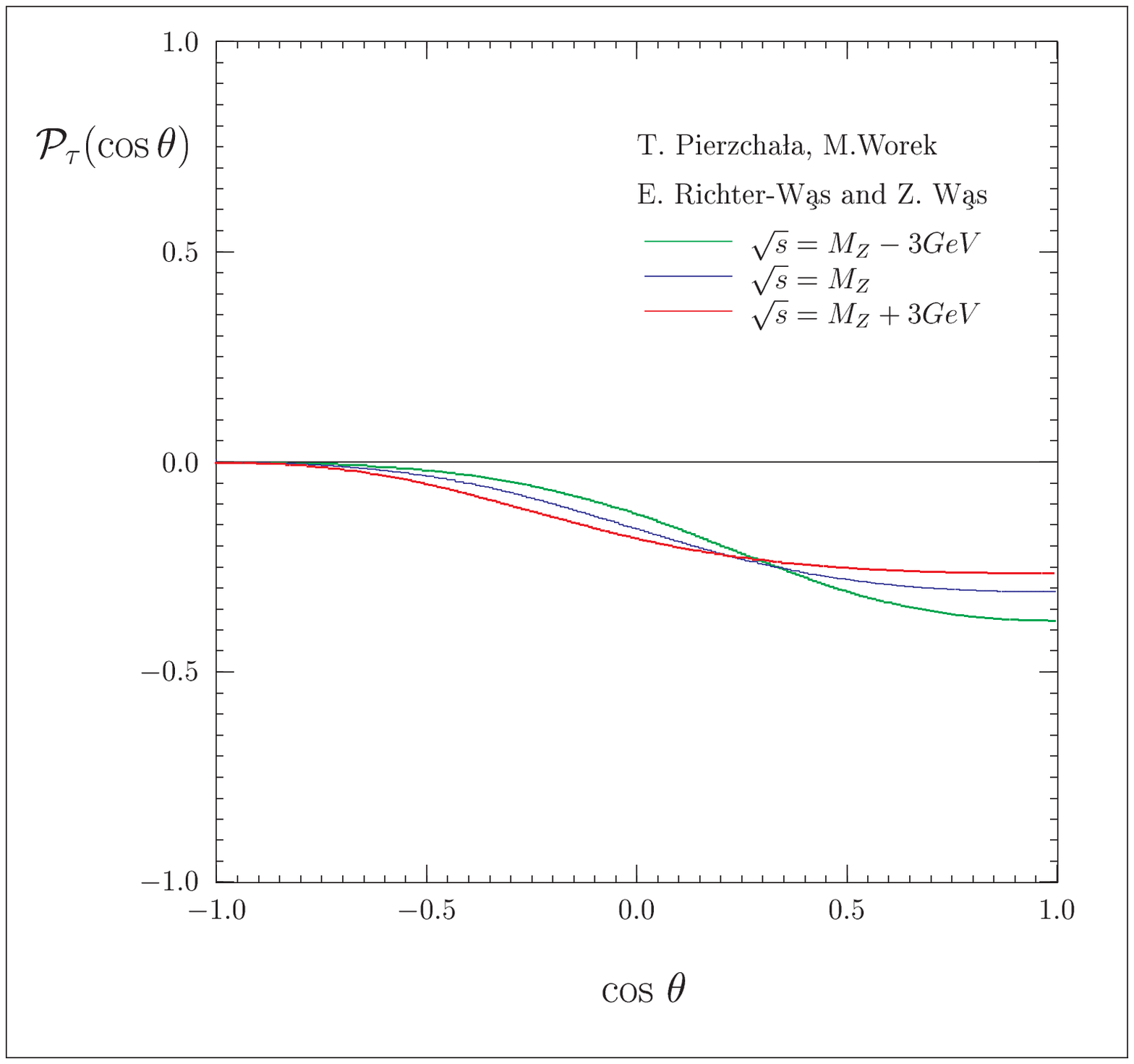,width=80mm,height=90mm}}}
\put(-20,  245){\makebox(0,0)[lb]{\epsfig{file=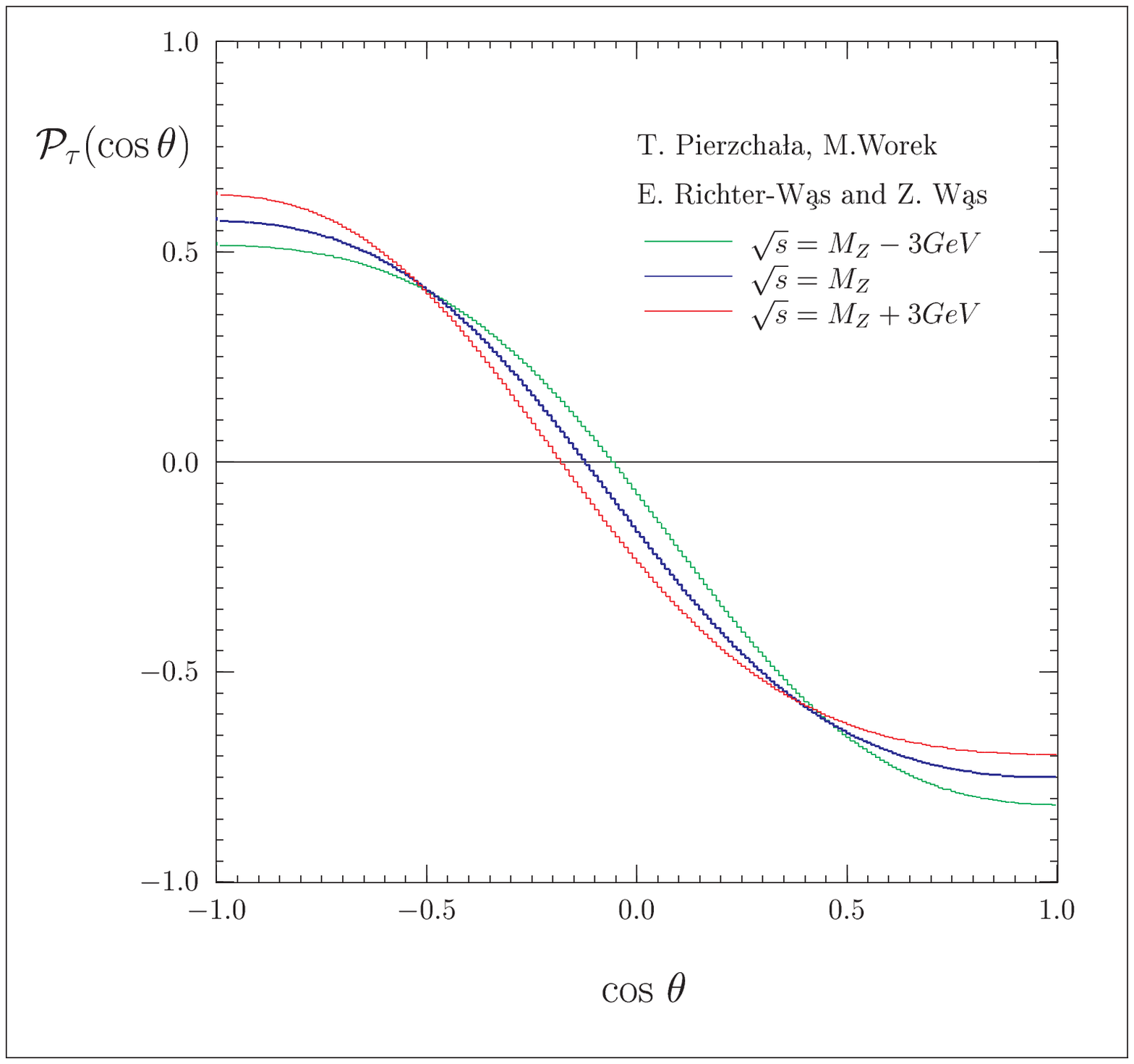,width=80mm,height=90mm}}}
\put(-20, -320){\makebox(0,0)[lb]{\epsfig{file=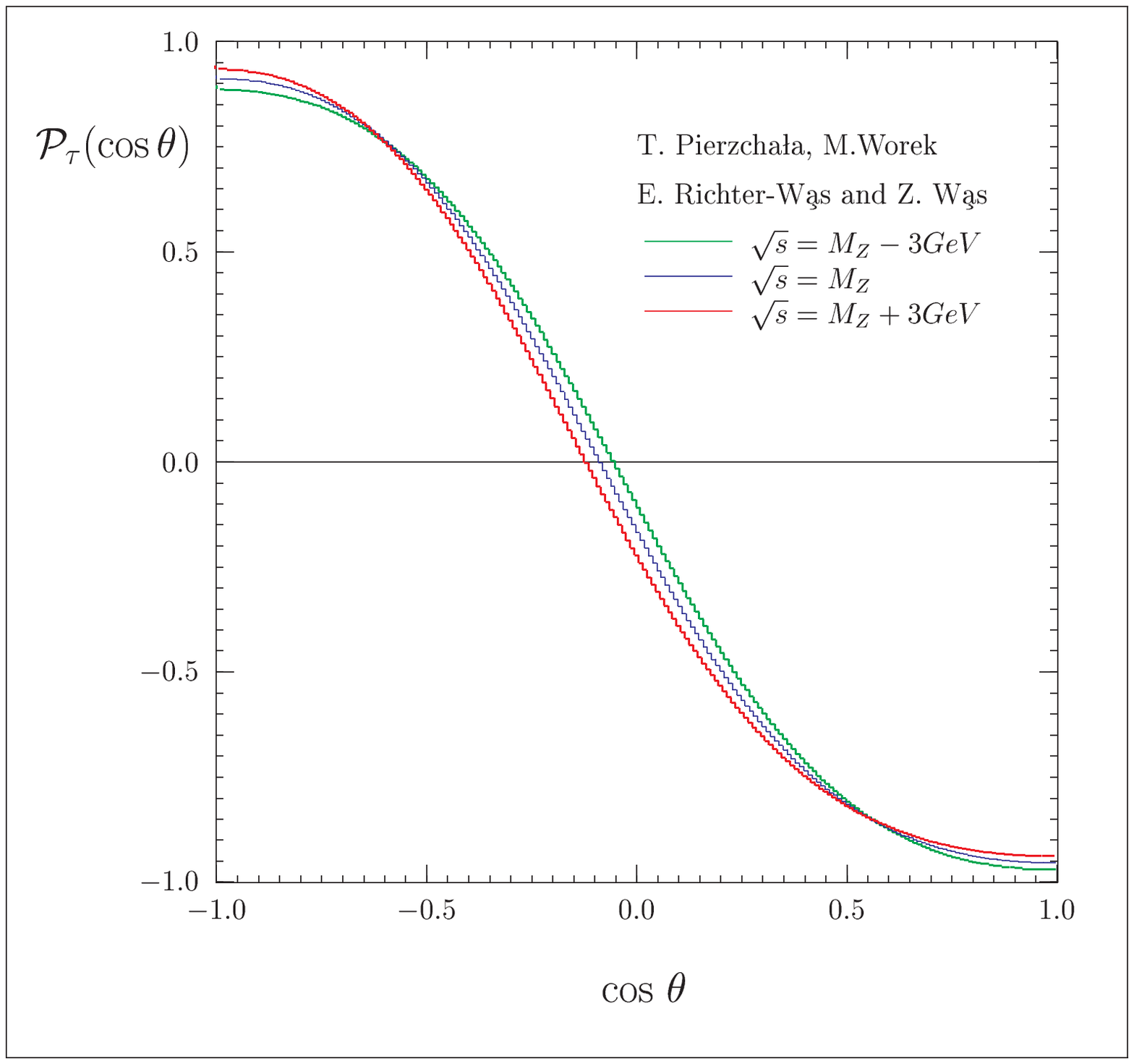,width=80mm,height=90mm}}}
\end{picture}
\caption{\small\sf Tests of the {\tt TAUOLA} universal interface. 
  The $\tau$ lepton  polarization as a function of $\cos\theta$. We have used 
$\sqrt{s}=M_{Z}$, $M_{Z}$ $\pm$ 3 GeV.
}
\label{fig:ifi-second}
\end{figure}


\begin{figure}[!ht]
\centering
\setlength{\unitlength}{0.1mm}
\begin{picture}(800,1200)
\put( 375,750){\makebox(0,0)[b]{\large }}
\put(1225,750){\makebox(0,0)[b]{\large }}
\put(-20, -320){\makebox(0,0)[lb]{\epsfig{file=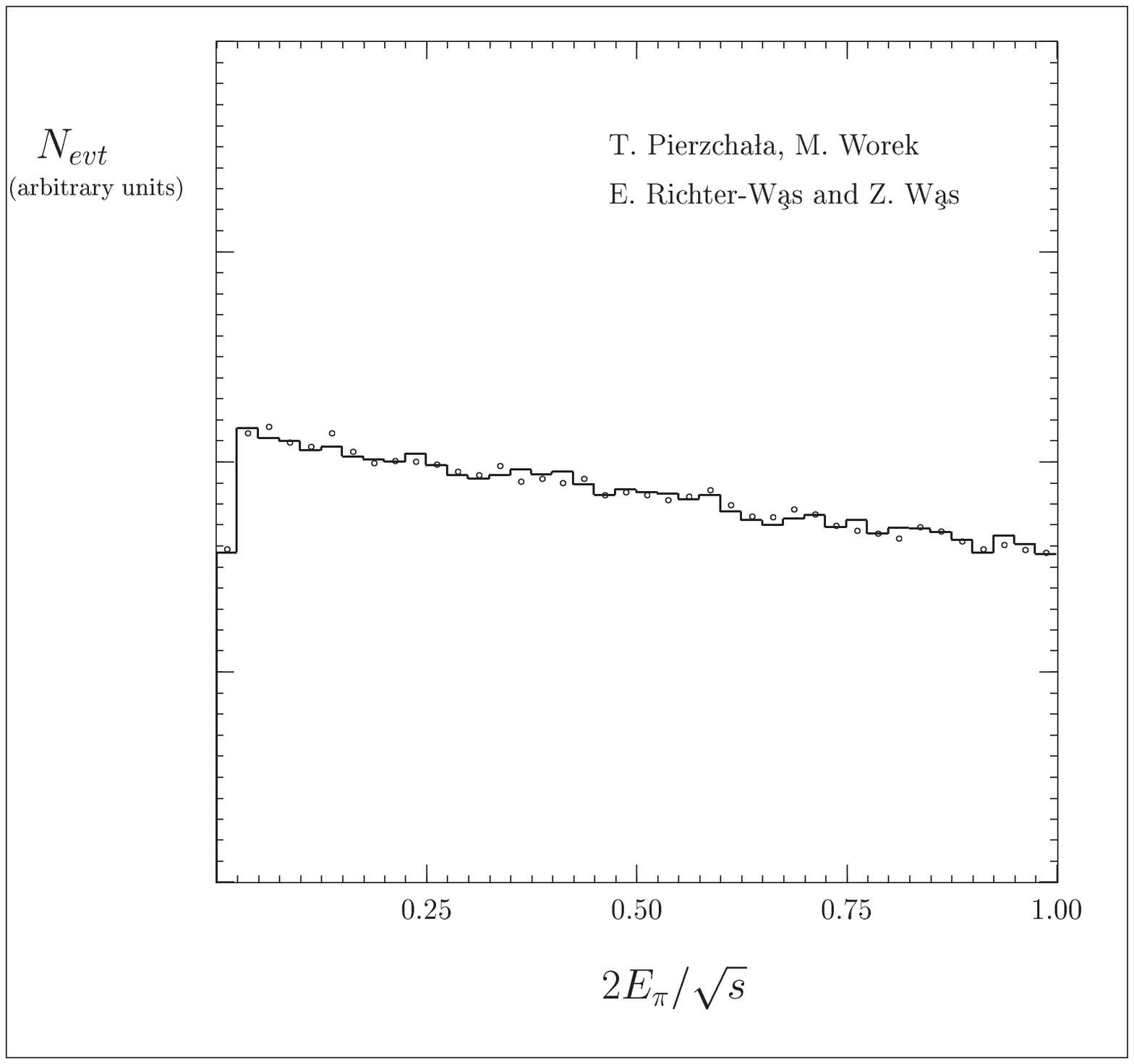,width=80mm,height=100mm}}}
\put(-20, 280){\makebox(0,0)[lb]{\epsfig{file=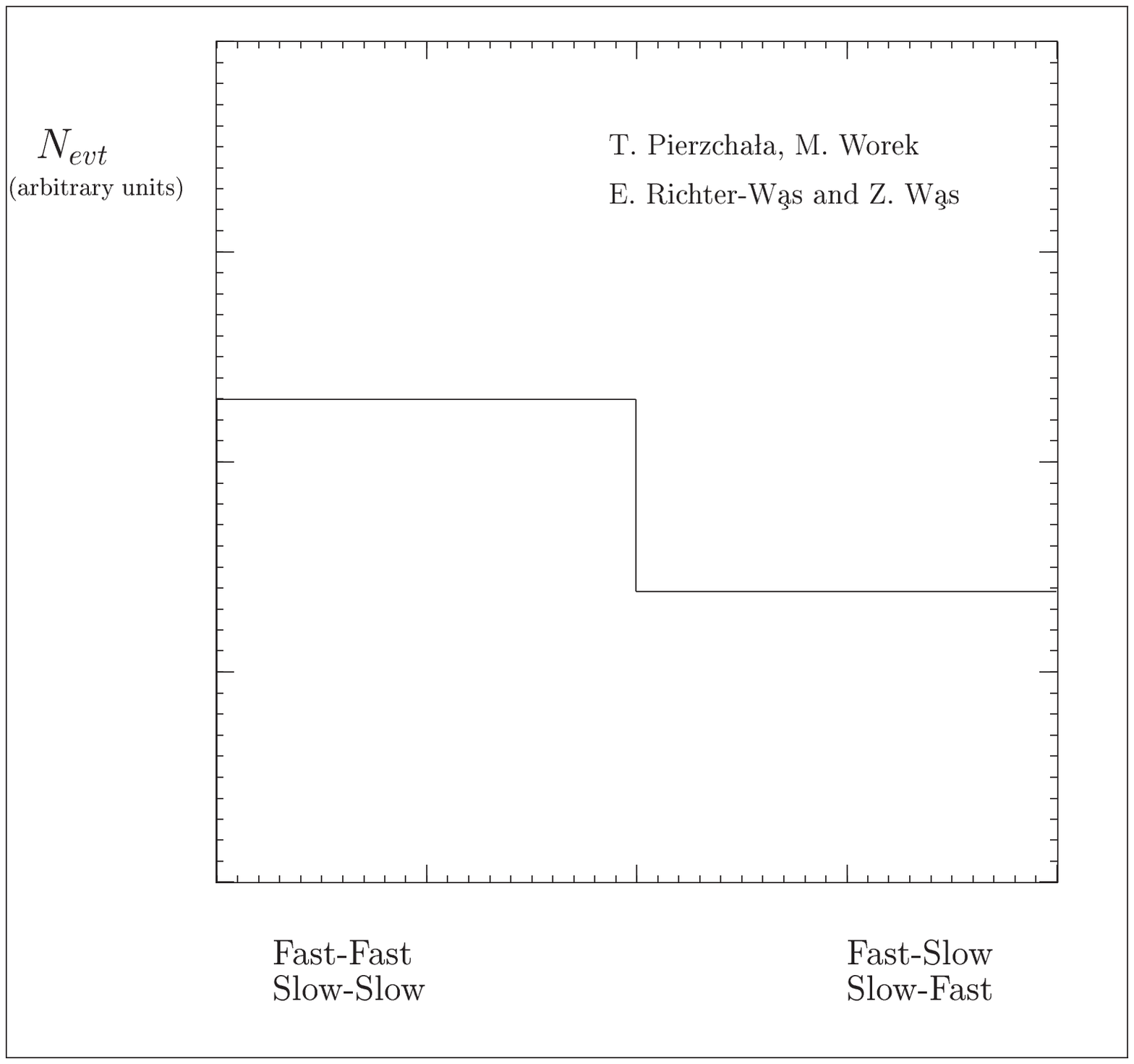,width=80mm,height=100mm}}}
\end{picture}
\caption[]{\small\sf Tests of the {\tt TAUOLA} universal interface, 
$\sqrt{s} \simeq M_{Z}$, spin effects included.\\
{\bf Lower plot:} The $\pi$ energy spectrum. \\
{\bf Upper plot:} $\pi^{+} \pi^{-}$ Energy-Energy correlations. \\
Both plots are for energies calculated in the $Z$ rest-frame.
}
\label{NuFig}
\end{figure}

\section{The $\tau$ pair production in $e^+e^-$ collisions}

Until 1999 the two widely used programs for the $\tau$-pair production
in $e^+e^-$ collisions were {\tt KORALB} \cite{koralb:1985} and {\tt KORALZ}
 \cite{koralz4:1994}.
The first of the two programs was used mainly at lower energies where its
full treatment of spin and $\tau$-mass effects was more important than
its limited, first-order only, treatment of radiative corrections.
{\tt KORALZ} was more appropriate for the higher energy zone, especially on the 
peak of the $Z$ 
resonance, where the ultra-relativistic approximation of the spin effects
was sufficient, effects of the higher order QED corrections were important,
and, thanks to the $Z$ life-time, the interference of initial and final state 
bremsstrahlung was strongly suppressed. This second assumption 
turned out not to withstand the numerical requirements of the LEP2 physics see eg.
\cite{Kobel:2000aw}. Also,  it is  not enough for  
requirements of the future high energy Linear Colliders requirements such as 
TESLA \cite{Brinkmann:2000jb}.

In the recent years, a new Monte Carlo
program, based on exponentiation performed at the spin amplitude level, was
developed \cite{kkcpc:1999}. Thanks to the applied technique, it can provide
predictions at the precision level of a few permille, for centre-of-mass energies 
from $\tau$ production threshold
up to the energies of the linear collider range.  So far, precision tests were
performed for the centre-of-mass energies corresponding to LEP2. For other
energies, points such as  matching virtual pair corrections with simulations
of 4-fermion final states \cite{Grunewald:2000ju}, photonic vacuum polarization effects
at low energies, Coulomb interactions, mass terms in some virtual corrections, 
etc. have to be discussed before precision at the permille level
can be granted. For more details and comparisons between {\tt KKMC, KORALZ} and 
{\tt KORALB},
see table 1 and ref. \cite{kkcpc:1999}.
\begin{table}
\begin{center}
\setlength{\unitlength}{1mm}
\begin{picture}(100,60)
\put(-2,0){\makebox(0,0)[lb]{\epsfig{file=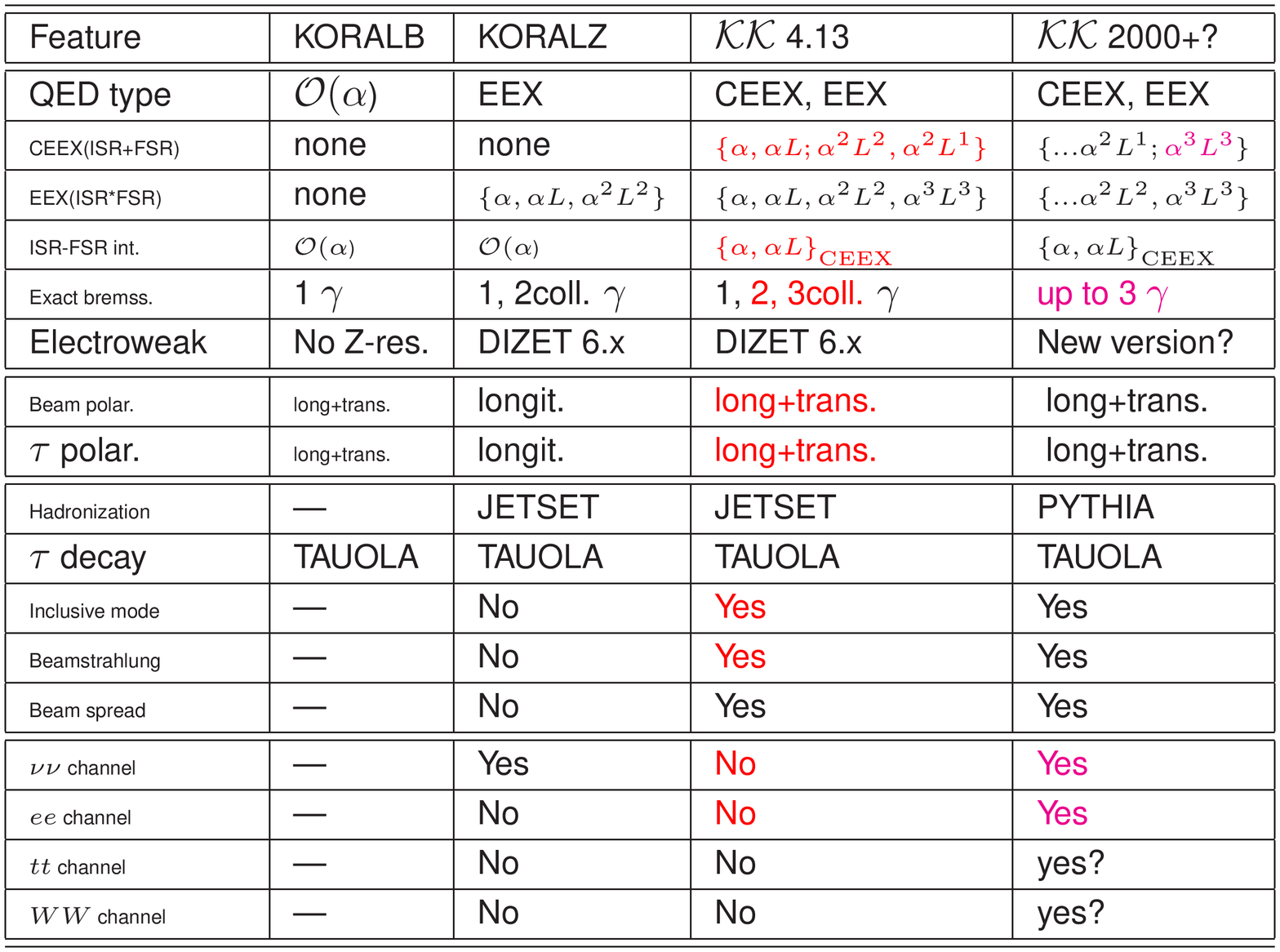,width=85mm,height=62mm}}}
\end{picture}
\\
{\small\sf {\bf Table 1:} Comparison of the main features for the {\tt KORALB, KORALZ,}
and {\tt KKMC} programs, see ref. \cite{kkcpc:1999} for more details.}
\end{center}
\end{table}

\section{Summary and future possibilities}

The status of the programs for the production of $\tau$ leptons was reviewed.
The high precision Monte Carlo program {\tt KKMC} for $\tau$-pair production 
in $e^+e^-$ collisions 
was presented. The general purpose interface, with the partial treatment of spin effects,
 of {\tt TAUOLA}  to any production  program was also reviewed.
At present, in its full content, it is available from the authors upon individual request only.

Distinct versions of the {\tt TAUOLA} library for $\tau$ lepton decay, and of
{\tt PHOTOS} for radiative corrections in decays, are now in use. 
The versions differ either
due to physics initialization or requirements of the interfaces. 
We have presented the system for creating the required version of the {\tt PHOTOS}
and {\tt TAUOLA} packages from the single  master copies. The master copies 
are kept
in relatively compact and clear  form, without code duplications and with the help
of the {\tt cpp} pre-compiler.

We expect that this organization, on top of the practical goals of the every day applications,
may serve also as a step to
 develop the packages (without loss of their physics content) into new internal 
architectures.
Some additional experience was collected in  \cite{MsCGolonka} in context of {\tt OO C++}.
We have found that the language translation for 
the given program version is relatively easy. On the contrary,
the question of project continuity into further upgrades
motivated by the physics needs to be thought over carefully. Matching the 
programming styles, e.g. of the {\tt OO C++}
experts, with the strategies of testing the numerical correctness of
consecutive versions is a rather crucial issue which has to be
addressed.  Tools and methods embodied 
in  {\tt FORTRAN} survive such translations with difficulty.

The successful  strategy will probably  require simultaneous fluency, at a certain step of 
the project development,
in  {\tt FORTRAN},  {\tt OO} languages {\it and} the 
physics content of the project, by all involved  persons. Platform independent 
tools for mixing code in  {\tt FORTRAN} and {\tt OO} languages will  be of great help.

\vskip 3 mm
\centerline{ \bf Acknowledgements}
\vskip 3 mm

Author is grateful to ALEPH, CLEO, DELPHI and OPAL collaborations
for providing their appropriate versions of the whole or parts of TAUOLA 
initialization. Useful discussions and suggestions from B. Bloch, S. Jadach,
J. H. K\"uhn, B.F.L Ward and 
A. Weinstein are acknowledged, as well as 
contributions from  P. Golonka, T. Pierzchala, E. Richter-Was and
M. Worek. The
warm hospitality of the Zurich ETH L3 group at the final stage of the project
completion is also acknowledged.
\bibliographystyle{utphys_spires}

\begingroup\begin{thebibliography}{10}

\bibitem{Jadach:1990mz}
S.~Jadach, J.~H. Kuhn, and Z.~Was, {\em Comput. Phys. Commun.} {\bf 64} (1990)
275.

\bibitem{Jezabek:1991qp}
M.~Jezabek, Z.~Was, S.~Jadach, and J.~H. Kuhn, {\em Comput. Phys. Commun.} {\bf
  70} (1992)
69.

\bibitem{Jadach:1993hs}
S.~Jadach, Z.~Was, R.~Decker, and J.~H. Kuhn, {\em Comput. Phys. Commun.} {\bf
  76} (1993)
361--380.

\bibitem{Golonka:2000iu}
P.~Golonka, E.~Richter-Was, and Z.~Was,
\href{http://www.arXiv.org/abs/hep-ph/0009302}{{\tt hep-ph/0009302}}.

\bibitem{Barberio:1990ms}
E.~Barberio, B.~van Eijk, and Z.~Was, {\em Comput. Phys. Commun.} {\bf 66}
  (1991)
115.

\bibitem{Barberio:1994qi}
E.~Barberio and Z.~Was, {\em Comput. Phys. Commun.} {\bf 79} (1994)
291--308.

\bibitem{aleph}
{ALEPH} Collaboration, B.~Bloch, to contact send a mail to \\
  Brigitte.Bloch-Devaux@cern.ch.

\bibitem{cleo}
{CLEO} Collaboration, A.~Weinstein, see \\
  http://www.cithep.caltech.edu/\~{}ajw/ korb$\_$doc.html\#{}files.

\bibitem{kkcpc:1999}
S.~Jadach, Z.~W\c{a}s, and B.~F.~L. Ward, {\em Comput. Phys. Commun.} {\bf 130}
  (2000) 260, Up to date source available from http://home.cern.ch/jadach/.

\bibitem{jetset6.3:1987}
T.~Sjostrand and M.~Bengtsson, {\em Comput. Phys. Commun.} {\bf 43} (1987)
367.

\bibitem{Abbiendi:1999cq}
{OPAL} Collaboration, G.~Abbiendi {\em et al.}, {\em Eur. Phys. J.} {\bf C13}
  (2000) 197,
\href{http://www.arXiv.org/abs/hep-ex/9908013}{{\tt hep-ex/9908013}}.

\bibitem{Abreu:1998cn}
{DELPHI} Collaboration, P.~Abreu {\em et al.}, {\em Phys. Lett.} {\bf B426}
  (1998)
411--427.

\bibitem{Gosia}
T.~Pierzchala, E.~Richter-Was, Z.~Was, and M.~Worek, unpublished, in
  preparation.

\bibitem{koralz4:1994}
S.~Jadach, B.~F.~L. Ward, and Z.~W\c{a}s, {\em Comput. Phys. Commun.} {\bf 79}
  (1994) 503.

\bibitem{was:1987}
Z.~W\c{a}s, {\em Acta Phys. Polon.} {\bf B18} (1987) 1099.

\bibitem{koralb:1985}
S.~Jadach and Z.~W\c{a}s, {\em Comput. Phys. Commun.} {\bf 36} (1985) 191.

\bibitem{Kobel:2000aw}
{Two Fermion Working Group} Collaboration, M.~Kobel {\em et al.},
\href{http://www.arXiv.org/abs/hep-ph/0007180}{{\tt hep-ph/0007180}}.

\bibitem{Brinkmann:2000jb}
{TESLA} Collaboration, R.~Brinkmann, {\em eConf} {\bf C000821} (2000)
MO202.

\bibitem{Grunewald:2000ju}
M.~W. Grunewald {\em et al.},
\href{http://www.arXiv.org/abs/hep-ph/0005309}{{\tt hep-ph/0005309}}.

\bibitem{MsCGolonka}
P.~Golonka, MSc thesis written under supervision of Z. Was, P. Golonka home
  page at {\tt http://lhotse.ifj.edu.pl/\~{}piters}.

\end{thebibliography}\endgroup

\providecommand{\href}[2]{#2}


\end{document}